\documentclass[apj]{emulateapj}
\usepackage{apjfonts}

\submitted{To appear in ApJ}

\renewcommand{\vec}[1]{\mbox{\boldmath $\displaystyle #1$}}
\newcommand{\grad}{\vec{\nabla}}

\newcommand{\vdot}{\vec{\cdot}}
\newcommand{\vcross}{\vec{\times}}
\newcommand{\divr}{\grad\vdot\,}
\newcommand{\curl}{\grad\vcross\,}

\begin{document}
 
\title{Latitudinal Shear Instabilities during Type I X-ray Bursts}
 
\author{Andrew Cumming}
\affil{Physics Department, McGill University, 3600 rue University, Montreal, QC, H3A 2T8, Canada; cumming@physics.mcgill.ca}

\begin{abstract}
Coherent oscillations have been observed during thermonuclear (Type I) X-ray bursts from 14 accreting neutron stars in low mass X-ray binaries, providing important information about their spin frequencies. However, the origin of the brightness asymmetry on the neutron star surface producing these oscillations is still not understood. We study the stability of a zonal shearing flow on the neutron star surface using a shallow water model. We show that differential rotation of $\gtrsim 2$\% between pole and equator, with the equator spinning faster than the poles, is unstable to hydrodynamic shear instabilities. The unstable eigenmodes have low azimuthal wavenumber $m$, wave speeds 1 or 2\% below the equatorial spin rate, and e-folding times $\lesssim 1\ {\rm s}$. Instability is related to low frequency buoyantly driven r-modes that have a mode frequency within the range of rotation frequencies in the differentially rotating shell. In addition, a modified Rayleigh's criterion based on potential vorticity rather than vorticity must be satisfied. The eigenfunctions of  the unstable modes have largest amplitudes nearer to the rotational pole than the equator. We discuss the implications for burst oscillations. Growth of shear instabilities may explain the brightness asymmetry in the tail of X-ray bursts. However, some fine tuning of the level of differential rotation and a spin frequency near $300\ {\rm Hz}$ are required in order for the fastest growing mode to have $m=1$. If shear instabilities are to operate during a burst, temperature contrasts of $\gtrsim 30$\% across the star must be created during ignition and spreading of the flash. The frequency drift may be related to the non-linear development of the instability as well as the cooling of the burning layer. The properties and observability of the unstable modes vary strongly with neutron star rotation frequency.
\end{abstract}


\section{Introduction}

Type I X-ray bursts are thermonuclear flashes on the surface of accreting neutron stars in low mass X-ray binaries (LMXBs) (see Lewin, van Paradijs, \& Taam 1993, 1995; Strohmayer \& Bildsten 2003 for reviews). Nearly-coherent oscillations with frequencies in the range $45$--$620\ {\rm Hz}$ have been seen during Type I X-ray bursts from 14 sources (Strohmayer et al.~1996; Strohmayer \& Bildsten 2003 and references therein; Strohmayer et al.~2003; Kaaret et al.~2003; Villarreal \& Strohmayer 2004). It is thought that rotational modulation of a brightness asymmetry on the neutron star surface produces an oscillation in X-ray intensity at the spin frequency of the neutron star (e.g.~Strohmayer \& Markwardt 1999). This has provided spin measurements for these LMXB neutron stars, confirming their rapid spin, as expected if they are the progenitors of the millisecond radio pulsars (Bhattacharya 1995).

The observational properties of burst oscillations have been intensively studied. The oscillations are sinusoidal (Strohmayer \& Markwardt 1999; Muno et al.~2000); they drift upwards in frequency by a few Hz during the burst, reaching an asymptotic frequency that is stable to a few tenths of a percent on timescales of years (Strohmayer et al.~1998b; Muno et al.~2002a); they are observed when the source is accreting at higher accretion rates rather than lower (Muno et al.~2000; van Straaten et al.~2001; Franco 2001; Muno, Galloway, \& Chakrabarty 2004); they are seen in the burst tail as well as the rise, with amplitudes typically 5\% in the tail (Muno, \"Ozel, \& Chakrabarty 2002) and much larger during the burst rise (Strohmayer, Zhang, \& Swank 1997b; Strohmayer et al.~1998a); their amplitude grows with photon energy (Muno, \"Ozel, \& Chakrabarty 2003). That the oscillations occur at the stellar spin frequency has been confirmed by detection of burst oscillations from two of the accreting ms X-ray pulsars (Chakrabarty et al.~2003; Strohmayer et al.~2003). Burst oscillations were also seen during the hours-long superburst from 4U~1636-54 (Strohmayer \& Markwardt 2002).

The physics of the burst oscillations has, however, remained unsolved. Initial work concentrated on the upwards frequency drift, which Strohmayer et al.~(1997a) suggested was related to angular momentum conservation in the burning layer, which expands outwards when heated and slowly contracts during the cooling tail of the burst. However, detailed models of the vertical expansion (Cumming \& Bildsten 2000; Cumming et al.~2002) showed that the expected spin down from this effect was a factor of 2--3 smaller than observed. Spitkovsky et al.~(2002) studied the ignition and propagation of a hotspot during the burst rise. They argued that rapid rotation is crucial because it allows the horizontal pressure gradients associated with a localized hot spot to be supported by hurricane-like winds. They showed that this leads to a new mode of burning front propagation in which the ageostrophic fluid flow associated with expansion of the hotspot brings in fresh fuel, allowing the burning to propagate, and spread around the star on a timescale of a second or less.

The picture of a spreading hotspot during the burst rise is appealing, and consistent with the large amplitudes observed (Strohmayer et al.~1997b). It does not however explain why burst oscillations are observed in the cooling tail, when the whole surface has ignited. There have been two mechanisms suggested that might cause asymmetries in the burst tail. Spitkovsky et al.~(2002) pointed out that the spreading burning front leaves behind temperature variations across the surface of the star, with associated zonal flows. They speculated that these flows were unstable to vortex formation, and pictured a vortex trapped in the zonal flow somewhat analogous to Jupiter's Great Red Spot. 

The second suggestion is that oscillation modes are excited in the burning layer, and that the frequency drift is due to a changing mode frequency as the burning layer cools (Heyl 2004; Lee 2004; Piro \& Bildsten 2005; see also McDermott \& Taam 1987; Strohmayer \& Lee 1996; Piro \& Bildsten 2004). In particular, Heyl (2004) identified low azimuthal wavenumber, buoyancy-driven r-modes as promising candidates, since they drift backwards in the rotating frame, and have a relatively large latitude coverage. However, the frequency of the lowest radial order mode in the burning layer is too large to explain the observed drifts, and so there remained a question as to whether a mode with the correct frequency exists, and how it is excited.  Piro \& Bildsten (2005) answer the first question by proposing that the surface wave associated with the burning layer transitions to a crustal interface mode during the burst tail, by an avoided crossing as the layer cools. Because the crustal interface mode has a frequency that is independent of the temperature in the burning layer, this terminates the frequency drift at a frequency that depends only on properties of the crust/ocean interface.

In this paper, we connect the ideas of unstable zonal flows and oscillation modes by studying the linear stability of shearing zonal flows during the tail of a Type I X-ray burst. We show that for differential rotation between pole and equator of $\gtrsim 2$\%, there are linearly unstable modes with properties that may be relevant for understanding burst oscillations. The properties of the unstable modes are closely related to the r-modes of a rigidly rotating layer. The unstable modes have low azimuthal wavenumbers $m$, growth rates $\lesssim 1\ {\rm s}$, and wave speeds of order $1$\% smaller than the equatorial rotation rate. We start in \S 2 by discussing the zonal flows expected during the tail of a burst, and then in \S 3 describe calculations of the linear stability of a quasi-geostrophic shallow water model of the burning layer. We then discuss the implications for our understanding of burst oscillations in \S 4. In the Appendix, solutions to the full set of shallow water equations are presented, and compared with the quasi-geostrophic solutions calculated in \S 3.

%-----------------------------------------------------------------------------------------------------------------------------------

\section{Zonal flows during the tail of a burst}\label{sec:zonal}

Spitkovsky et al.~(2002) emphasized that a zonal flow naturally arises as the burning front spreads around the star. If the local cooling time is the same at each point, the part of the surface that ignited first will be cooler by a factor of $\approx t_{\rm spread}/t_{\rm cool}$ than the part that ignited last, where $t_{\rm spread}$ is the time for the burning front to spread across the surface, and $t_{\rm cool}$ is the local cooling time. Because the spin period of the star is much shorter than either of these timescales, Coriolis forces act to produce a zonal flow which balances the lateral pressure gradient associated with the differential cooling.

We start by estimating the amount of differential rotation expected. We model the burning layer using the shallow water equations (e.g. Houghton 1986; Pedlosky 1987), in which the detailed vertical structure of the layer is neglected. We work in the rotating frame. The momentum equation is
\begin{equation}\label{eq:momentum}
{d\vec{u}\over dt}+2\vec{\Omega}\vcross\vec{u}=-g\grad H,
\end{equation}
and the continuity equation is
\begin{equation}\label{eq:continuity}
{dH\over dt}=-H\divr\vec{u},
\end{equation}
where $\vec{u}=\hat e_\theta u_\theta(\theta,\phi)+\hat e_\phi u_\phi(\theta,\phi)$ is the horizontal fluid velocity, $H$ is the pressure scale height of the layer, $g=GM/R^2$ is the surface gravity and $\Omega$ is the neutron star spin frequency. Throughout this paper, $\theta$ is chosen such that $\theta=0$ corresponds to the pole, and $\theta=90^\circ$ the equator. The advective derivative $d/dt=\partial/\partial t+\vec{v}\vdot\grad$ involves horizontal velocities only, and also only Coriolis forces arising from horizontal motions are included in equation (\ref{eq:momentum}). In addition, we assume that the burning layer is spherical, and neglect distortions due to the centrifugal force.

We assume a steady state flow with azimuthal velocity $\vec{u}=u_\phi\hat e_\phi$, where we write $u_\phi(\theta)=\Delta\Omega(\theta)r\sin\theta$. The force balance is
\begin{equation}
2\Omega\cos\theta u_\phi  = {g\over r}{\partial H\over\partial\theta},
\end{equation}
or in terms of $\mu=\cos\theta$, 
\begin{equation}\label{eq:geo}
{\partial H\over\partial \mu}=-{2\Omega r^2\over g}\mu\Delta\Omega(\mu).
\end{equation}
We will consider in this paper the simplest shear profile symmetric about the equator,
\begin{equation}\label{eq:uphi}
\Delta\Omega=-s\mu^2\Omega,
\end{equation}
in which case integrating equation (\ref{eq:geo}) gives
\begin{equation}\label{eq:H}
H(\mu)=H_0\left(1+{sF\over 8}\mu^4\right),
\end{equation}
where $H_0$ is the scale height at the equator. The dimensionless parameter $F$ is
\begin{equation}\label{eq:F}
F={(2\Omega r)^2\over g H_0}=71\left({f_s\over 300\ {\rm Hz}}\right)^2 {R_6^2\over H_3}\left({2\over g_{14}}\right),
\end{equation}
where $R=10\ R_6\ {\rm km}$ and $g=g_{14}\ 10^{14}\ {\rm cm\ s^{-2}}$. Another way to write this is $F=(R/R_D)^2$, where in terms of the Brunt-V\"ais\"al\"a frequency $N$, $R_D\approx H(N/\Omega)\sim 1\ {\rm km}$ is the Rossby deformation radius, the horizontal scale on which rotational effects are important (Pedlosky 1987).

In equation (\ref{eq:F}), we write the pressure scale height $H$ in units of $10^3\ {\rm cm}$. At the high temperatures achieved during the burst, the ideal gas equation of state is appropriate, and radiation pressure may also play a role at the peak temperature (Cumming \& Bildsten 2000). Therefore, we estimate 
\begin{equation}\label{eq:scaleheight}
H={P\over \rho g}\approx {k_BT\over \beta\mu_m m_pg}=416\ {\rm cm}\ {T_9\over \mu_m\beta}\left({g_{14}\over 2}\right)^{-1},
\end{equation}
where $\mu_m$ is the mean molecular weight (we use a subscript $m$ to avoid confusion with $\mu=\cos\theta$), and $\beta$ is the ratio of gas to total pressure. For typical fluxes and base temperatures appropriate for the peak of a burst, Cumming \& Bildsten found $\beta\approx 0.5$, with $\beta$ rapidly approaching unity as the layer cools. Then,
\begin{equation}
F=171{\mu_m\beta\over T_9}\left({f_s\over 300\ {\rm Hz}}\right)^2R_6^2,
\end{equation}
where $T_9$ is the temperature in units of $10^9\ {\rm K}$.

Rewriting equation (\ref{eq:H}), the differential rotation between the pole and equator is 
\begin{equation}
s=\left({2H\over R}\right)\left({\Delta H\over H}\right)\left({\Omega_K\over\Omega}\right)^2,
\end{equation}
where $\Omega_K^2=GM/R^3=g/R$ (this is a similar result to Spitkovsky et al.~2002, eq.~[74]). This gives 
\begin{equation}
s=0.11\ \left({\Delta H\over H}\right)\left({f_s\over 300\ {\rm Hz}}\right)^{-2} {H_3\over R_6^2}\left({g_{14}\over 2}\right).
\end{equation}
If $s>0$, the equator is rotating faster than the poles, and the fluid piles up at the poles, $\Delta H>0$. 

If we follow Spitkovsky et al.~(2002) and take $\Delta H/H\approx t_{\rm spread}/t_{\rm cool}$, we find $s\approx 1$\%$\ (t_{\rm spread}/1\ {\rm s})(10\ {\rm s}/t_{\rm cool})$. The direction of the differential rotation will depend on the details of ignition and spreading of the burning. Spitkovsky et al.~(2002) argue that ignition starts at the equator, in which case at any given time as the burst cools the poles will be higher pressure relative to the equator. This implies $s>0$, or more rapid rotation at the equator than the poles. In fact, this case is more favorable for shear instabilities. We will see in \S 3 that instability sets in for much smaller levels of differential rotation if the equator is rotating more rapidly than the poles.

%-----------------------------------------------------------------------------------------------------------------------------------

\section{Lateral shear instabilities in the shallow water model}

\subsection{Perturbation equations with the geostrophic approximation}
\label{sec:geostrophic}

We now perturb the shallow water equations around the background zonal flow described in \S \ref{sec:zonal}. To simplify the calculation, and to highlight the physics of the instability, we will assume that the perturbed fluid velocity is in geostrophic balance. This approximation must in fact break down at the equator, where the vertical component of the rotation vector vanishes. However, away from the equator, we expect the approximation to be a good one. In the Appendix, we calculate the r-modes of the full set of shallow water equations, and show that they compare well with the geostrophic solutions except for a small region near the equator.

The assumption of geostrophic balance for the perturbations allows us to work with the vorticity equation only. Taking the curl of equation (\ref{eq:momentum}), and using equation (\ref{eq:continuity}), gives the vorticity equation for shallow water,
\begin{equation}
{d\over dt}\left({\omega_r+2\Omega\mu\over H}\right)=0,
\end{equation}
where $\omega_r=(\curl\vec{u})_r$ is the radial component of fluid vorticity in the rotating frame, $2\Omega\mu$ is the radial component of vorticity coming from the overall rotation, and $H$ is the thickness of the layer. For $H$ constant, this is simply a statement of conservation of vorticity; with a variable $H$, the potential vorticity $(\omega_r+2\Omega\mu)/H$ is conserved. The extra physics included in potential vorticity conservation (rather than ordinary vorticity conservation) is vortex stretching: as a column of fluid is stretched vertically, its vorticity changes to satisfy overall angular momentum conservation (e.g., Houghton 1986; Pedlosky 1987).

The vorticity of the background state is
\begin{equation}
\Omega_r=-(d/d\mu)\left[(\Omega+\Delta\Omega)(1-\mu^2)\right],
\end{equation}
which is $\Omega_r=2\Omega\mu$ for rigid rotation, and $\Omega_r=2\Omega\mu(1+s)-4\Omega s\mu^3$ for the rotation law of equation (\ref{eq:uphi}). Perturbing the vorticity equation then gives to linear order in the perturbations
\begin{eqnarray}
\left({\partial\over\partial t}+\Delta\Omega{\partial\over\partial\phi}\right)\omega_r&+&{\delta u_\theta\over r}{\partial \Omega_r\over\partial \theta}\\
&=&{\Omega_r\over H}\left({\partial\over\partial t}+\Delta\Omega{\partial\over\partial\phi}\right)h+{\Omega_r\over H}{\delta u_\theta\over r}{\partial H\over\partial \theta},\nonumber
\end{eqnarray}
where $H$ and $\Omega_r$ are the height and vorticity of the background, and $h$ and $\omega_r$ are the perturbations to the height and vorticity, and the perturbation to the fluid velocity is $\delta u_\phi\hat e_\phi+\delta u_\theta\hat e_\theta$.

Assuming that the perturbations are in geostrophic balance  provides a relation between the fluid velocities and $h$. We first write the perturbed velocity in terms of a stream function $\psi$, such that
\begin{equation}\label{eq:stream}
\delta u_\theta={1\over r\sin\theta}{\partial \psi\over\partial\phi};\hspace{1cm}
\delta u_\phi=-{1\over r}{\partial \psi\over\partial\theta},
\end{equation}
which automatically enforces $\divr\vec{u}=0$. Then $\omega_r=-{\mathcal L}\psi/r^2$, where
\begin{equation}\label{eq:legendre}
{\mathcal L}={d\over d\mu}\left[(1-\mu^2){d\over d\mu}\right]-{m^2\over 1-\mu^2}
\end{equation}
is the Legendre operator, and $m$ is the azimuthal wavenumber, i.e.~we have assumed a $\phi$ dependence $\propto e^{im\phi}$. We then use the $\phi$ component of the momentum equations under the assumption of geostrophic balance, $2\Omega\cos\theta u_\theta=({-g/r\sin\theta})({\partial h/\partial\phi})$, to derive a relation between $\psi$ and $h$,
\begin{equation}\label{eq:heightgeo}
h=-\left({2\Omega\mu\over g}\right)\psi.
\end{equation}
It is important to note that this relation is not exact, because it only approximately satifies the $\theta$ component of the momentum equations. The approximation is that geostrophic balance holds at the equator, which it does not. We also note here that this approximation removes gravity waves from the system.

We now look for solutions $\psi\propto \exp(-im\tilde\omega t+im\phi)$, where $\tilde\omega=\omega+i\sigma$ is the complex mode frequency. Also, we write the perturbation equation in dimensionless variables, with unit of time $\Omega^{-1}=0.53\ {\rm ms}\ (f_s/300\ {\rm Hz})^{-1}$, and velocity $\Omega R$. The final perturbation equation is
\begin{equation}\label{eq:perturb}
(\tilde\omega-\Delta\Omega){\mathcal L}\psi - \psi{d\Omega_r\over d\mu}- {\tilde\omega\mu\psi F\Omega_r\over 2H}=0,
\end{equation}
with ${\mathcal L}$ given by equation (\ref{eq:legendre}). With appropriate boundary conditions, this specifies an eigenvalue problem for the complex frequency $\tilde\omega$.

We solve the eigenvalue problem with a shooting method. We reduce equation (\ref{eq:perturb}) to two first order  differential equations for $\psi$ and $\Phi=(1-\mu^2)(d\psi/d\mu)$. Then, following Dikpati \& Gilman (1999) and Bildsten, Ushomirsky, \& Cutler (1996), we remove the singularity at $\mu=1$ by defining $f$ and $g$ such that $\psi=(1-\mu^2)^{\left|m\right|/2}f$ and $\Phi=(1-\mu^2)^{\left|m\right|/2}g$, which also enforces the boundary condition $\psi=0$ at the pole. The resulting equations are
\begin{equation}
(1-\mu^2){df\over d\mu}=g+\mu\left|m\right|f
\end{equation}
and
\begin{equation}
{dg\over d\mu}={\mu\left|m\right|g+m^2f\over 1-\mu^2}-{f\over\Delta\Omega-\tilde\omega}\left[\Omega_r^\prime +{\tilde\omega\mu F\Omega_r\over 2H}\right].
\end{equation}
We start integrating a small distance from the pole, by setting $f=1$ at $\mu=1-\epsilon$ and $g=-\mu\left|m\right|f$. For our choice of rotation law, the equations are symmetric under the change $\mu\rightarrow -\mu$. Given this, we integrate to the equator ($\mu=0$) and look for solutions which are either symmetric or antisymmetric about the equator (either $g=0$ or $f=0$ at $\mu=0$). Since the eigenvalue $\omega+i\sigma$ is complex, we integrate both real and imaginary parts of $f$ and $g$, and perform a two-dimensional search over $\omega$ and $\sigma$ using Newton's method.

%-----------------------------------------------------------------------------------------------------------------------------------

\begin{figure}
\epsscale{1.0}
\plotone{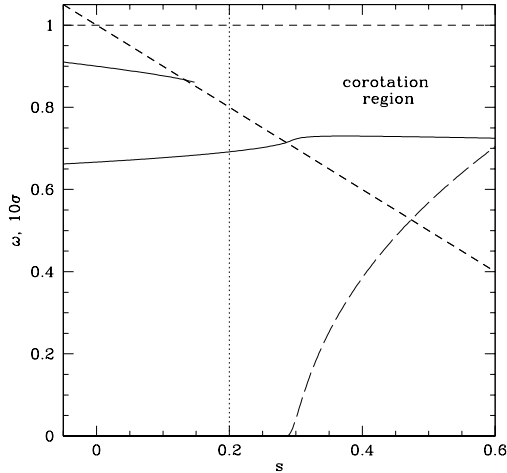}
\caption{Wave speeds and growth rates as a function of the differential rotation $s$ for 2D modes ($F\ll 1$). The background vorticity profile satisfies Rayleigh's condition for instability for $s>0.2$, or to the right of the dotted line. The dashed lines demark the ``corotation region'' in which a critical level exists in the flow. The two solid lines are for modes which are the $l=2, m=1$ and $l=4, m=3$ r-modes at zero differential rotation. The $l=4$ mode approaches the boundary of the corotation region and stops. The $l=2$ mode enters the corotation region and becomes unstable. The long-dashed curve shows the growth rate (multiplied by a factor of 10).\label{fig:om_hydro}}
\end{figure}

\subsection{Shear instabilities for a two-dimensional flow}
\label{sec:2D}

We present the results of the stability calculation in \S \ref{sec:results}. First, it is useful to summarize the results of stability of latitudinal shear in a two-dimensional spherical shell. In the shallow water model, the limit of a 2D shell is equivalent to holding $H$ constant, or setting $F=0$. This case has been studied by Watson (1981), Dziembowski \& Kosovichev (1987), Charbonneau, Dikpati, \& Gilman (1999), Watts et al.~(2003), Watts, Andersson, \& Williams (2004), and in the weakly non-linear regime by Garaud (2001). Much of this previous work was motivated by trying to understand the rotation profile of the Sun, in particular the role of shear instabilities in the solar tachocline. These results can be understood in a simple way which is helpful when discussing the shallow water instability. Dynamical instability is closely connected with the existence of a critical level at which the real part of the mode frequency is equal to the local rotation rate, as found for shear instabilities in differentially rotating disks (e.g.~Papaloizou \& Pringle 1985) (also, the existence of a critical level leads to a complex mode spectrum, including a continuous spectrum of neutral modes, see e.g.~Watts et al.~2003). 

\subsubsection{Rayleigh's criterion}

Watson (1981) derived necessary conditions for instability starting with equation (\ref{eq:perturb}) with $F=0$. Multiplying by $\psi^\star/(\tilde\omega-\Delta\Omega)$, integrating over $\mu$, and taking the imaginary part leads to the requirement that $d\Omega_r/d\mu$ must change sign somewhere in the domain for instability. This is the equivalent of Rayleigh's criterion for instability of parallel shear flows (e.g. Drazin \& Reid 1981), that there be an inflexion point in the velocity profile\footnote{Note that there is an additional necessary condition for instability, Fj\o rtoft's criterion, that is always satisfied for our choice of rotation law.}. For the rotation law of equation (\ref{eq:uphi}), $d\Omega_r/d\mu=2(1+s-6s\mu^2)$, which gives the necessary condition for instability $s>1/5$ or $s<-1$. Note that instability is possible for much smaller differential rotation when the equator is rotating faster than the pole, as opposed to the pole rotating faster than the equator. For this reason, we will concentrate on the $s>0$ case in this paper.

\begin{figure}
\epsscale{1.0}
\plotone{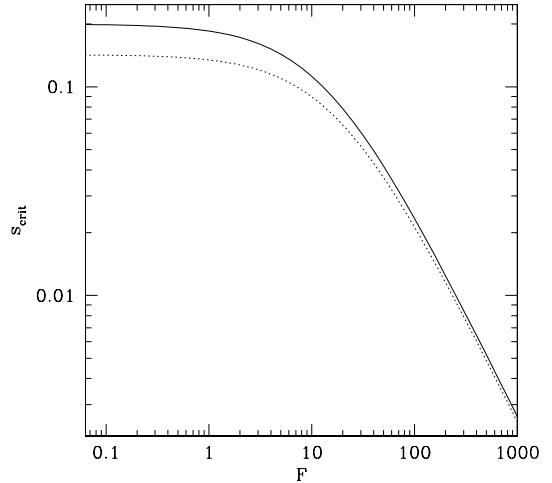}
\caption{The critical value of $s$ needed to satisfy Rayleigh's criterion for the shallow water equations as a function of $F$. Solid line is for the rotation law $\Omega=1-s\mu^2$, dotted line is for $\Omega=1-(s/2)(\mu^2+\mu^4)$. For $F\ll 1$, $s_{\rm crit}=1/5$ and $1/7$ respectively for these rotation laws. For $F\gg 1$, $s_{\rm crit}\approx 8/3F$ and $48/19F$ respectively.\label{fig:crit2}}
\end{figure}

It has been pointed out in studies of the Sun's rotation profile that introducing a $\mu^4$ term in the rotation law leads to instability at a lower level of differential rotation than with a $\mu^2$ term only (Charbonneau, Dikpati, \& Gilman 1999). This can be understood from Rayleigh's criterion. For example, the rotation law $\Omega=1-(s/2)(\mu^2+\mu^4)$ considered by Garaud (2001) requires $s>1/7=0.14$ for Rayleigh's criterion to be satisfied, rather than $s>1/5$ as previously, allowing instability to set in at a lower differential rotation. Interestingly, this is not the case for $s<0$, for which the requirement is still $s<-1$ even when a $\mu^4$ term is included.

\subsubsection{The r-modes, critical levels and instability}

A rigidly-rotating 2D shell supports a set of r-modes, with frequency in the rotating frame
\begin{equation}
m\omega=-{2m\Omega\over l(l+1)},
\end{equation}
and eigenfunctions $P_l^m(\mu)$, with $m\leq l$ (to see this, set $F=0$, $\Delta\Omega=0$, and $d\Omega_r/d\mu=2\Omega$ in eq.~[\ref{eq:perturb}]). The inertial frame frequency is $m(\Omega+\omega)$. For $m=1$, the modes with $l$ odd (even) are symmetric (antisymmetric) about the equator, and opposite for $m=2$. As described by Watts et al.~(2003), the instability can be understood in terms of the evolution of these modes as the level of differential rotation is increased. In Figure \ref{fig:om_hydro} we show two examples of the evolution of the mode frequency with increasing $s$, focusing on the $s>0$ case. The critical value for Rayleigh's criterion $s=1/5$ is indicated by the dotted line. As the level of differential rotation is increased, the onset of instability is associated with the development of a ``critical level'' in the flow, at which the mode frequency is equal to the local rotation frequency, or $\omega=\Omega(\mu)$. This occurs for points within the wedge defined by the lines $\omega=1$ and $\omega=1-s$, shown as short-dashed lines in Figure \ref{fig:om_hydro}. 

The unstable modes are those which, as differential rotation increases, satisfy Rayleigh's criterion at the point at which they first develop a critical level (Watts et al.~2003). The solid curves in Figure \ref{fig:om_hydro} show the evolution of the $l=4$, $m=3$ and $l=2$, $m=1$ r-mode frequencies as differential rotation is introduced. The $l=4$ mode approaches the corotation region when the background vorticity profile has not yet met Rayleigh's criterion (to the left of the dotted line), and does not cross the corotation boundary. The $l=2$ mode enters the corotation region with Rayleigh's criterion satisfied (to the right of the dotted line), and becomes an unstable mode. The long-dashed curve in Figure \ref{fig:om_hydro} shows the growth rate of the mode, which increases with increasing $s$.

We find instability for $s\geq 0.29$, in agreement with Watson (1981). The reason for this is that the unstable modes have $\omega\approx 0.7$ because they are related to the $l=2$ r-mode, and so $s\gtrsim 0.3$ is needed in order for a corotation point to exist in the flow. This differential rotation profile has two modes that are unstable, the $l=2, m=1$ antisymmetric mode, and the $l=2, m=2$ symmetric mode. Note that modes with $m\geq 3$ have $l\geq 3$, giving them frequencies too large to satisfy the instability requirements. The only unstable modes are $m=1$ or $m=2$ (see also Watson 1981 for a different argument as to why only low $m$ modes are unstable).

%-----------------------------------------------------------------------------------------------------------------------------------

\subsection{Shear instability in the shallow water model}
\label{sec:results}

In this section, we describe the results of our stability calculation for $F>0$. Before presenting the unstable eigenmodes, we discuss the general nature of the instability by deriving a generalization of Rayleigh's criterion for the shallow water model, and the spectrum of r-modes.

\begin{figure}
\epsscale{1.0}
\plotone{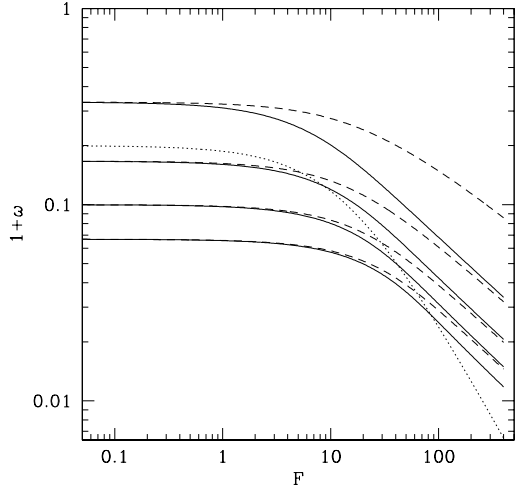}
\caption{The r-mode frequencies for $m=1$ (solid) and $m=2$ (dashed) modes with  $l=2,3,4,5$ as a function of $F$ for rigid rotation ($s=0$). For large $F$, the scaling is $\omega\propto F^{-1/2}$, as predicted by the asymptotic analysis of Longuet-Higgins (1968) (see eq.~[\ref{eq:asymptotic}]). The dotted curve shows the critical value of $s$ needed for Rayleigh's criterion to be satisified at a given value of $F$. Only modes which lie above the dotted line are expected to be unstable.
\label{fig:froude3}}
\end{figure}

\begin{figure}
\epsscale{1.0}
\plotone{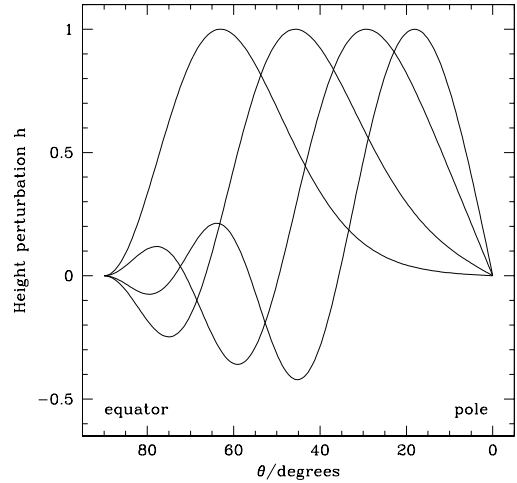}
\caption{Antisymmetric $m=1$ r-mode eigenfunctions, for $s=0$, $F=100$. We show the height perturbation $h$ in one hemisphere, normalized to be unity at the maximum value. These modes have $\nu=1,3,5$ and $7$. The frequencies are $-0.0682, -0.0312, -0.0211,$ and $-0.0160$. The predicted turning points for these modes are at $\mu=(2\nu+1)^{1/2}/F^{1/4}$ (see Appendix) which gives $57^\circ$, and $33^\circ$ for $\nu=1$ and $3$, in good agreement with the numerical solutions. The remaining two modes are able to propagate all the way to the pole.\label{fig:rmodes}}
\end{figure}

\begin{figure*}
\epsscale{1.0}
\plottwo{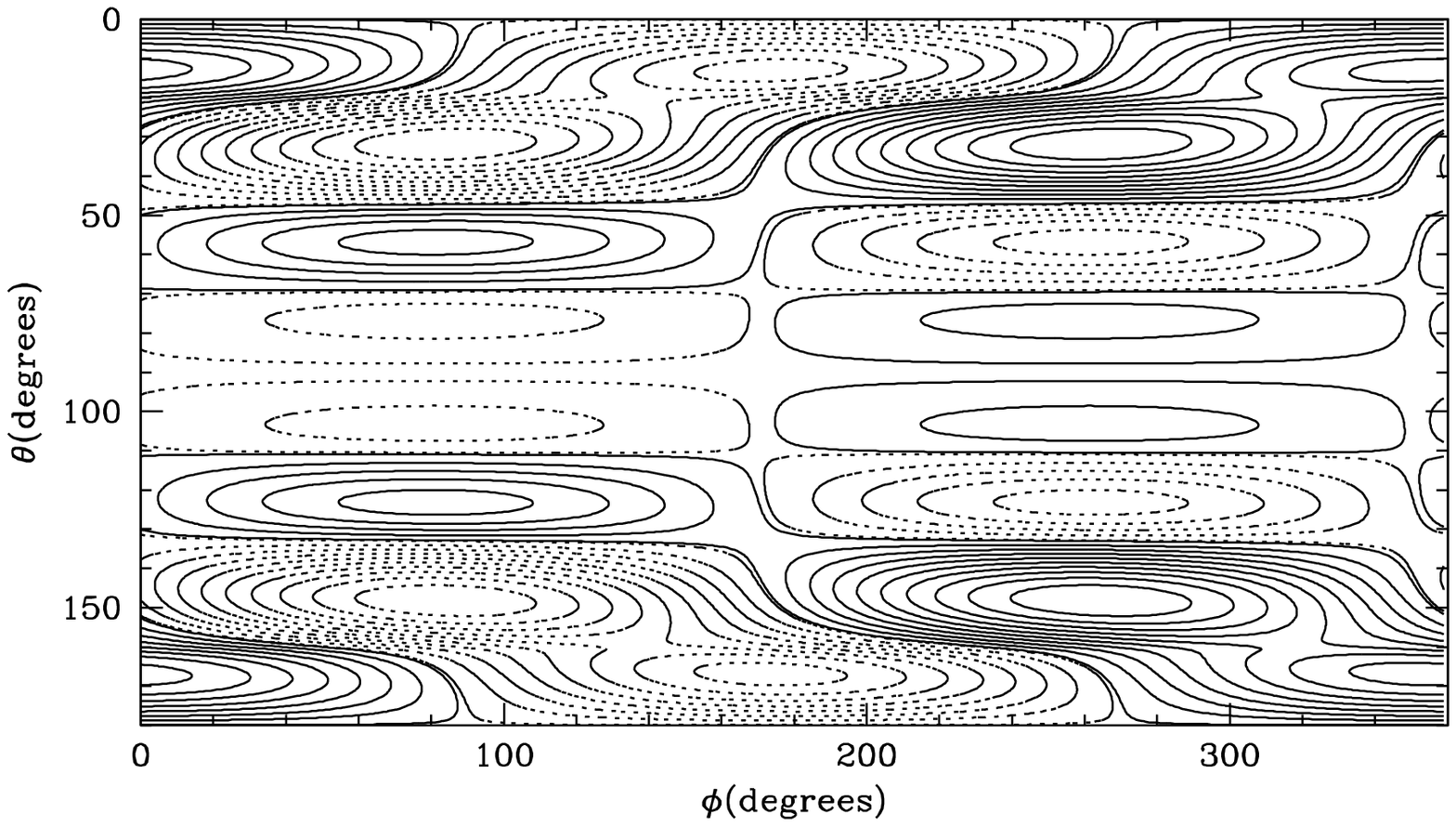}{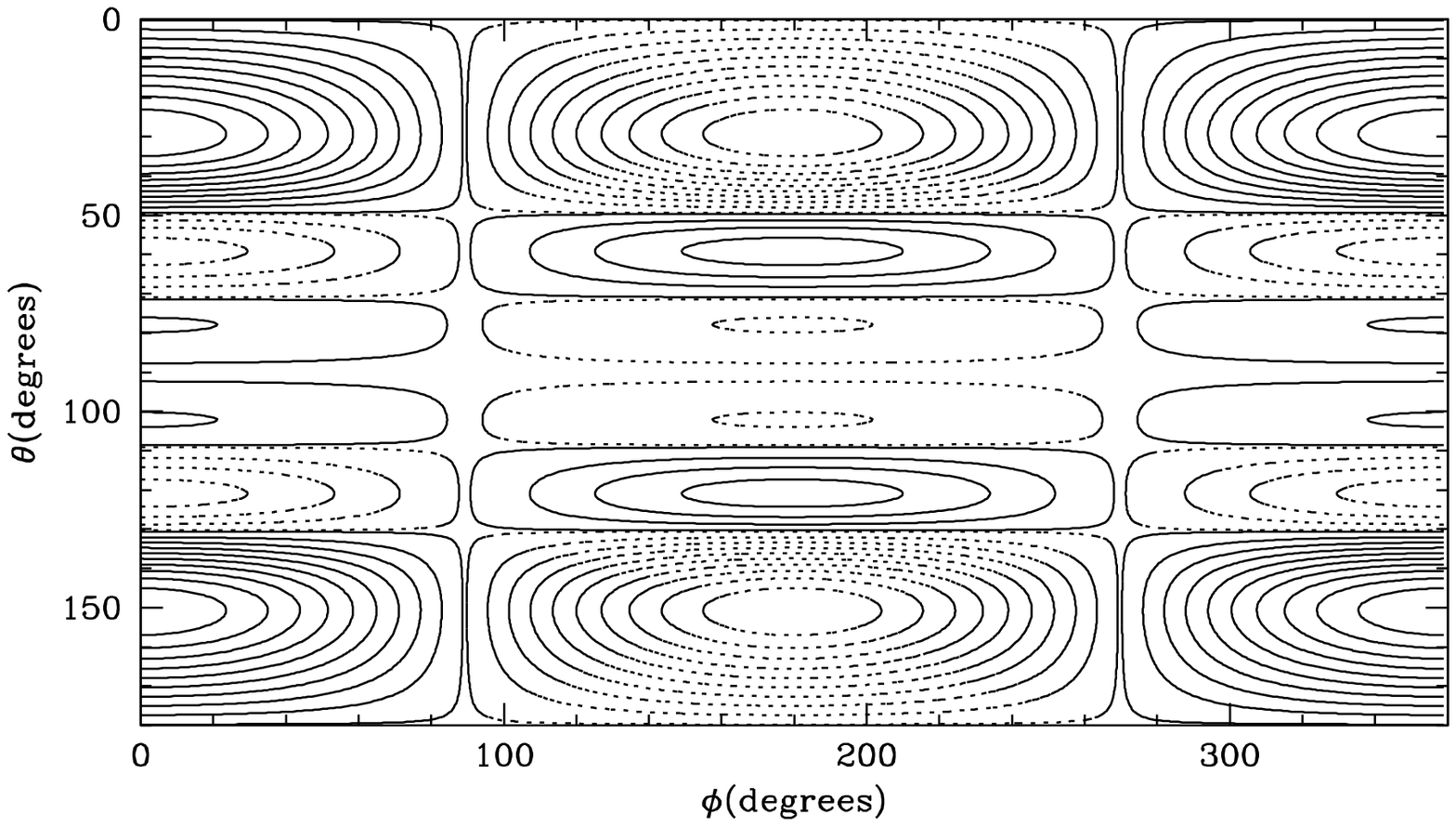}
\caption{(a) Height contours for the unstable mode shown in Figure \ref{fig:mode1}, which has $\omega=-0.0267$ ($\nu_{\rm eff}=3.2$). The height perturbation is strongly sheared at the critical level $\mu_{\rm crit}=(\omega/s)^{1/2}$, approximately 20 degrees from the pole. Solid contours are positive, dotted contours are negative. (b) For comparison, the $m=1$, $F=100$, $\nu=5$ r-mode for a rigidly rotating background. The mode frequency is $\omega=-0.0211$.\label{fig:mode1cont}}
\end{figure*}

\begin{figure*}
\epsscale{0.8}
\plottwo{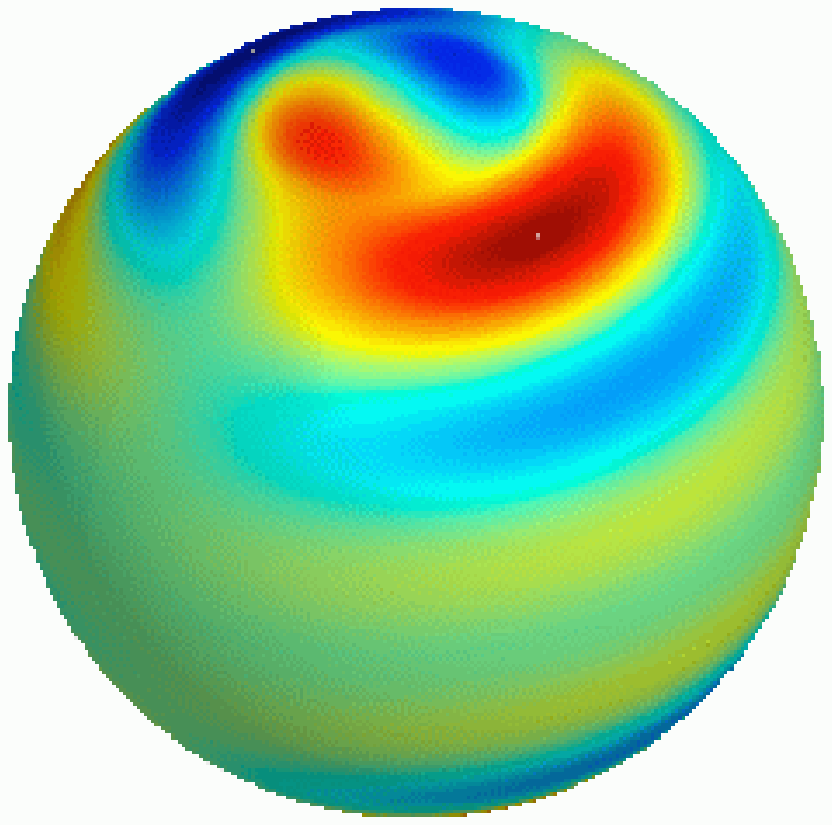}{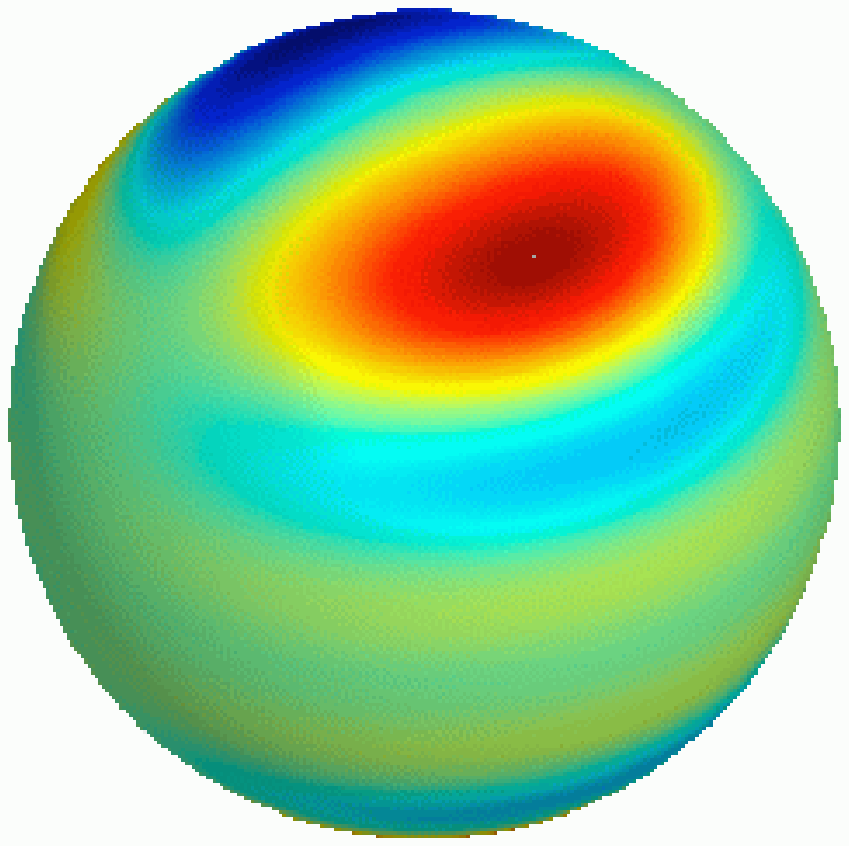}
\caption{As Figure \ref{fig:mode1cont}, but projected onto a sphere. Colored contours indicate the height perturbation.}
\end{figure*}

\subsubsection{Rayleigh's criterion}

To derive the equivalent of Rayleigh's criterion for the shallow water equations, we multiply equation (\ref{eq:perturb}) by $\psi^\ast/(\tilde\omega-\Delta\Omega)$, and integrate over $\mu$ from $\mu=-1$ to $\mu=+1$. Taking the imaginary part of the resulting equation gives
\begin{equation}
\sigma\int^{1}_{-1}d\mu\ {\left|\psi\right|^2\over \left|\tilde\omega-\Delta\Omega\right|^2}\left[
\Omega_r^\prime+{\mu F\Omega_r\Delta\Omega\over 2H}\right]=0.
\end{equation}
A non-zero growth rate $\sigma$ requires that the quantity in the square brackets change sign somewhere in the domain. Using equation (\ref{eq:geo}), we can write this factor as 
\begin{equation}
\Omega_r^\prime+{\mu F\Omega_r\Delta\Omega\over 2H} = H{d\over d\mu}\left({\Omega_r\over H}\right),
\end{equation}
which is the potential vorticity gradient. We see that as suggested by Dikpati \& Gilman (1999), in shallow water it is the potential vorticity gradient that must change sign in the domain, rather than the vorticity gradient, thereby including the extra physics of vortex stretching in Rayleigh's criterion.

We evaluate the critical amount of differential rotation $s_{\rm crit}$ needed for instability by setting $(d/d\mu)(\Omega_r/H)=0$ at $\mu=1$, because by inspection we have found that the critical level always appears at the pole as $s$ is increased beyond its critical value for our assumed rotation law. Using equations (\ref{eq:uphi}) and (\ref{eq:H}) for $\Delta\Omega$ and $H$, we find that $s_{\rm crit}$ is given by
\begin{equation}\label{eq:scrit}
F={8-40s_{\rm crit}\over s_{\rm crit}(3+s_{\rm crit})}
\end{equation}
or $s_{\rm crit}\approx 8/3F$ for large $F\gg 40/3$. 

We plot $s_{\rm crit}$ as a function of $F$ in Figure \ref{fig:crit2}. For $F\approx 100$, only $\approx 2$\% differential rotation between pole and equator is required for instability to be possible, an order of magnitude smaller than the 2D case. The dotted curve in Figure \ref{fig:crit2} shows the critical $s$ needed for instability for the rotation law $1-(s/2)(\mu^2+\mu^4)$. For this rotation law, we find $F=48(1-7s_{\rm crit})/(s_{\rm crit}(19+11s_{\rm crit}))$, giving a very similar critical value in the $F\gg 1$ limit, $s_{\rm crit}\approx 48/19F=2.53/F$ as compared to $8/3F=2.67/F$ previously. In fact, a general rotation law $\Delta\Omega=-s\mu^n$ has $s_{\rm crit}=(2/F)(n+2)/(n+1)$ for $F\gg 1$. The critical value $s_{\rm crit}F$ translates into a critical height contrast between equator and pole, using equation (\ref{eq:H}), of $\Delta H/H\gtrsim 1/3$. For the more general rotation law $\Delta\Omega\propto \mu^n$, this becomes $\Delta H/H\gtrsim 1/(n+1)$.

\subsubsection{The r-mode spectrum}

We next discuss the r-mode spectrum for rigid rotation ($s=0$). The effect of increasing $F$ is to reduce the rotating frame r-mode frequency (see also the calculations of Lee 2004). Figure \ref{fig:froude3} shows the r-mode frequency in the rotating frame, $1+\omega$, as a function of $F$ for different $m$ and $l$ values. On the left, the modes have $l=2,3,4$ and $5$ from top to bottom. On the right, for large $F$, modes with the same $l-m$ have almost the same frequencies. In fact, the r-modes of a thin, spherical shell were analysed in detail by Longuet-Higgins (1968), who derived an asymptotic form for the mode frequency in the limit $F\gg 1$,
\begin{equation}\label{eq:asymptotic}
m\omega=-{2m\Omega\over (2\nu+1)F^{1/2}}
\end{equation}
(eq.~[8.33] of Longuet-Higgins 1968), where $\nu=l-m$ is a new quantum number for large $F$. This compares well with our calculated mode frequencies (also, compare our Figure \ref{fig:froude3} with Figures 2 and 3 of Longuet-Higgins 1968). 
Later, we will use the quantity $\nu_{\rm eff}=((2/F^{1/2}\omega)-1)/2$ as a label for the modes. Using the definition of $F$, the frequency in the rotating frame in the limit $F\gg 1$ can be written as $m\omega=-m(\sqrt{gH}/R)/(2\nu+1)$, a direct multiple of the surface gravity wave frequency in a non-rotating star. 

The eigenfunctions for $F>0$ are known as Hough functions (Longuet-Higgins 1968), shown in Figure \ref{fig:rmodes} for antisymmetric modes with $\nu=1,3,5$ and $7$ and $F=100$.  Note that the number of latitudinal nodes in the height perturbation $h$ is $\nu-1$. Modes with $\nu$ odd are antisymmetric in $\psi$ about the equator, or symmetric in $h$. Modes with $\nu$ even are symmetric in $\psi$ about the equator, or antisymmetric in $h$. Note however, that in the geostrophic approximation, $h$ vanishes at the equator even for odd $\nu$ modes (such as those shown in Fig.~\ref{fig:rmodes}) because the Coriolis force vanishes there. This is not the case for solutions to the full shallow water equations (see Appendix).

Analytic solutions to the full shallow water equations in the limit $F\gg 1$ were derived by Longuet-Higgins (1968). In the Appendix, we derive a similar solution under the quasi-geostrophic assumption. In that case, the perturbation equation reduces to the equation for a simple harmonic oscillator in quantum mechanics. The eigenfunction is able to propagate  from the equator to a turning point at  $\mu=\sqrt{2\nu+1}/F^{1/4}=(2\nu+1)q^{-1/2}$, and then evanesces towards the pole. The maximum height displacement is close to the turning point. Therefore, while rapid rotation (large $F$) concentrates the eigenfunction towards the equator, this is less so for large $\nu$ values. For example, the $\nu=3$ mode with $F=100$ has a turning point at $33^\circ$ from the pole, and maximum amplitude at $45^\circ$ (Fig.~\ref{fig:rmodes}). Modes with $\nu>(F^{1/2}-1)/2$ can propagate all the way to the poles ($\nu\geq 5$ for $F=100$; $\nu\geq 8.2$ for $F=300$). The r-modes are much less confined than the g-modes, which have a turning point at $\mu=1/q$ (e.g.~Bildsten et al.~1996).

\begin{figure}
\epsscale{1.0}
\plotone{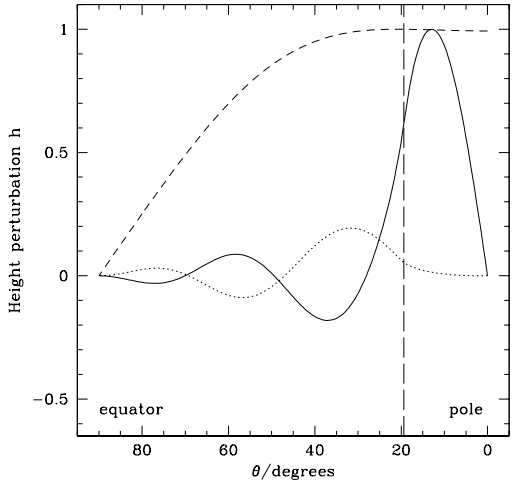}
\caption{An example of an unstable mode, the $m=1$ antisymmetric mode for $s=0.03$, $F=100$. The solid and dotted curves show the real and imaginary part of the height perturbation. The mode frequency is $\omega=-0.0267$, and growth time $1.87\ {\rm s}$. The critical level for this mode is shown by the vertical dashed line ($\mu_{\rm crit}=(\omega/s)^{1/2}$, approximately 20 degrees from the pole). The short dashed line shows the potential vorticity profile (normalized so that it's maximum value is unity). For this mode, $\nu_{\rm eff}=((2/\omega F^{1/2})-1)/2=3.2$.\label{fig:mode1}}
\end{figure}

 %-----------------------------------------------------------------------------------------------------------------------------------

\subsubsection{Unstable modes}

\begin{figure}
\epsscale{1.0}
\plotone{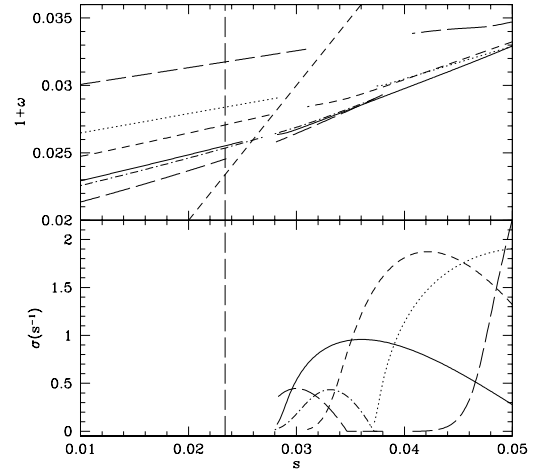}
\caption{Mode frequency and growth rate for a variety of modes with $F=100$. The modes are: $m=1$ antisymmetric (solid curves), $m=1$ symmetric (dotted), $m=2$ symmetric (short dashed), $m=2$ antisymmetric (long dashed), and $m=3$ symmetric (dot-dashed). The mode has a corotation point to the right of the short dashed straight line. Rayleigh's criterion is satisfied to the right of the vertical long-dashed line ($s_{\rm crit}=0.0234$ for $F=100$, from eq.~[\ref{eq:scrit}]). Note that the mode frequency does not include a factor of $m$. The growth rate does include the factor of $m$, and is given in units appropriate for a star with $f_s=300\ {\rm Hz}$ (it may be rescaled using $\sigma\propto f_s$).\label{fig:mode3}}
\end{figure}

The discussion so far suggests that shear instabilities will be important for the $\sim 1$\% levels of differential rotation expected during Type I X-ray bursts. First, we have shown that a much smaller level of differential rotation satisfies Rayleigh's criterion for large $F$ than for the 2D case. Secondly, the r-mode frequencies are much closer to the spin frequency than in the 2D case, so that it is possible for them to develop a critical level at such low levels of differential rotation. The dotted curve in Figure \ref{fig:froude3} shows the critical value of differential rotation $s_{\rm crit}$ needed for Rayleigh's criterion to be satisfied. For $F\ll 1$, only the $l=2$ mode lies above the dotted curve. We saw in \S \ref{sec:2D} that this mode (either with $m=1$ or $m=2$) is associated with the development of dynamical instability as it crosses into the corotation region. For the values of $F\gtrsim 100$ expected during Type I X-ray bursts, several modes have frequencies above the dotted curve, suggesting that dynamical instability is possible for large enough differential rotation that one of these modes develops a critical level\footnote{In Figure \ref{fig:froude3} we are comparing the r-mode frequency based on rigid rotation ($s=0$) with the critical value of $s$. This is reasonable because the r-mode frequencies do not change substantially as $s$ is increased from zero to the critical value (e.g.~see Fig.~\ref{fig:mode3}).}.

In fact, we do find unstable modes in this region. Figures \ref{fig:mode1cont} and \ref{fig:mode1} show an example of an unstable $m=1$ eigenmode for $F=100$ and $s=0.03$. This mode has a rotating frame frequency of $-0.0267$ (which gives $\nu_{\rm eff}=3.2$), and e-folding time $1.87\ {\rm s}$. In Figure \ref{fig:mode1}, we plot the real and imaginary parts of the height perturbation. The vertical dashed line shows the critical level, given by $-\omega=s\mu^2$, in this case at $20^\circ$ from the pole. The short-dashed curve shows the background potential vorticity profile. The mode frequency is such that the maximum in the potential vorticity profile (ie.~the latitude where $d(\Omega_r/H)/d\mu=0$) occurs close to (but not at) the critical level. This is a common feature of the unstable modes. The maximum amplitude of the unstable mode occurs polewards of the critical level, as found for 2D instabilities (Charbonneau et al.~1999). Figure \ref{fig:mode1cont} shows the perturbation height contours for the unstable mode and an r-mode for comparison. The height contours are strongly sheared at the critical level. As discussed by Dikpati \& Gilman (1999), this acts to transport angular momentum from the equator to the poles.

Figure \ref{fig:mode3} shows the mode frequencies and growth rates for $F=100$ as a function of $s$. From equation (\ref{eq:F}), we see that this value of $F$ is appropriate for the early stages of a burst on a star with $f_s=300\ {\rm Hz}$. To the right of the vertical long-dashed line ($s>s_{\rm crit}$), the background potential vorticity profile satisfies Rayleigh's criterion. To the right of the sloping short-dashed line, the modes have a critical level. At a given value of differential rotation larger than $s_{\rm crit}$, we find several unstable modes. In Figure \ref{fig:mode3}, we start from either $s=0$ or $s=0.05$, and work across the diagram, approaching the corotation boundary from either the left or right. At some point close to the corotation boundary, our code jumps across the boundary, moving from either an unstable mode to a stable one, or vice versa. Unlike the 2D case shown in Figure \ref{fig:om_hydro}, we have not been able to track a given mode across the boundary of the corotation region. We expect this is because of numerical difficulties as the growth rate approaches zero, and the critical level becomes singular. A more detailed study along the lines of Watts et al.~(2003) for the 2D case is needed to understand the behavior of the modes close to and at the corotation boundary. Nevertheless, we can see clearly in Figure \ref{fig:mode3} that there is a correspondence between r-modes outside the corotation region and unstable modes inside.

\begin{figure}
\epsscale{1.0}
\plotone{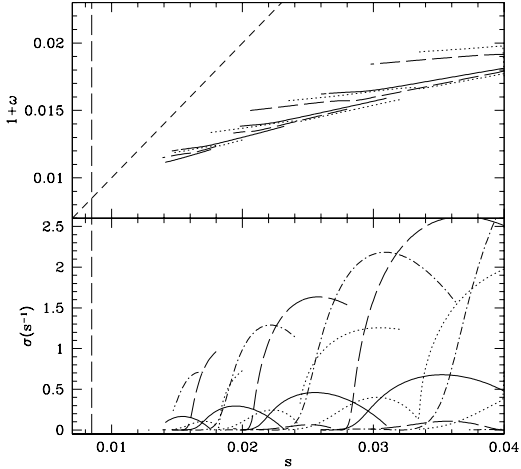}
\caption{As Figure \ref{fig:mode3}, but for $F=300$. In this plot for clarity we do not distinguish between symmetric and antisymmetric modes, and we do not show the stable r-modes (which are located to the left of the short-dashed straight line). Alternating peaks in the growth rate as $s$ increases correspond to alternating symmetry. We show $m=1$ (solid), $m=2$ (dotted), $m=3$ (short-dashed), and $m=4$ (dot-dashed) modes. For $F=300$, Rayleigh's criterion is satisfied for $s>s_{\rm crit}=0.00849$, indicated by the vertical long-dashed line.\label{fig:mode5}}
\end{figure}

\begin{figure}
\epsscale{1.0}
\plotone{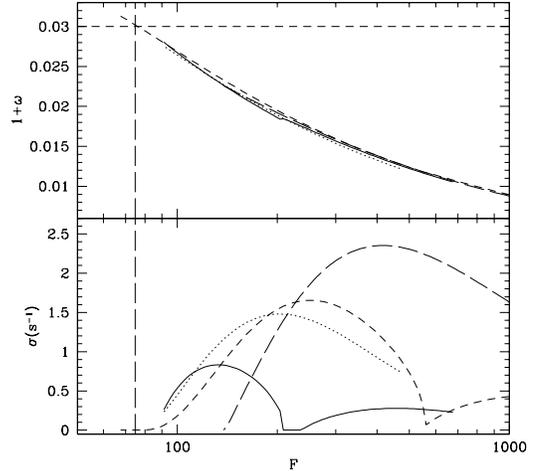}
\caption{Similar to Figure \ref{fig:mode3}, but now for a fixed value of $s=0.03$, the variation in mode frequency and growth rate with $F$. We show the $m=1$ antisymmetric (solid), $m=2$ antisymmetric (dotted), $m=3$ symmetric (short-dashed), and $m=4$ symmetric (long-dashed) modes. For clarity, we show either antisymmetric or symmetric for each value of $m$, choosing the symmetry which gives the largest growth rate in this range of $F$. The horizontal short-dashed line shows the mode frequency below which there is a critical level. The long dashed line shows the critical value of $F$ needed to satisfy Rayleigh's criterion at $s=0.03$ ($F_{\rm crit}=75$).\label{fig:mode4}}
\end{figure}

Figure \ref{fig:mode3} shows that as $s$ increases, the most unstable mode has larger $m$, larger growth rate, and the mode frequency in the rotating frame increases. For $s$ near the threshold, $s\approx 0.03$, the most unstable mode is antisymmetric and $m=1$. However, for $s\approx 0.04$, the most unstable mode is symmetric and $m=2$. The growth rates are of order $\sim 0.001 \Omega$, or $\sim 100$ rotation periods. For a star with $f_s=300\ {\rm Hz}$, this is $\sigma\sim 1\ {\rm s^{-1}}$. Figure \ref{fig:mode5} shows the unstable modes for $F=300$. As would be expected from Figure \ref{fig:froude3}, more unstable modes are found at this larger value of $F$, and we find instability for a smaller value of $s$. The most unstable modes in this case are those with $m=3$ or $m=4$, and they have larger growth rates than the $F=100$ case. The curves in Figure \ref{fig:mode4} terminate suddenly in many cases because we are not able to follow that particular mode any further, usually because the code switches to another unstable mode.

Figure \ref{fig:mode4} shows the variation with $F$ at a fixed value of $s=0.03$. The vertical dashed line shows the critical value of $F$ needed to satisfy Rayleigh's criterion; below the horizontal dashed line a mode has a critical level. As $F$ increases, the most unstable mode has larger $m$. We can understand this by inspecting the eigenfunction as $m$ increases. With increasing $F$ or $s$, the maximum in the potential vorticity moves towards the equator. Increasing $m$ provides a way to move the maximum of the eigenfunction in the same direction. To see this, note that the term involving $m$ in equation (\ref{eq:perturb}) (hidden inside the operator ${\mathcal L}$) is most important at the poles, where $1-\mu^2$ diverges. In a WKB sense, the effect is to increase the $k_\phi$ component of the horizontal wavevector ${\mathcal L}\sim k_\perp^2=k_\mu^2+k_\phi^2$. The mode amplitude increases more slowly away from the pole, and the maximum in the mode amplitude occurs closer to the equator.

Shear instabilities in the shallow water model were calculated by Dikpati \& Gilman (1999) for application to the Sun, using the full shallow water equations. Our results agree well with those of Dikpati \& Gilman (1999) for the values of $s\gtrsim 0.15$ they considered, and help to understand some of their results. In particular, they noted that the perturbed fluid velocities were in geostrophic balance except near the equator; this provides an additional justification for our ageostrophic approximation. They found that the $m=1$ and $m=2$ growth rates decreased for $F\gtrsim 400$ (or for their parameter $G=4/F\lesssim 0.01$), which is roughly where we find that the most unstable mode has $m>2$. Higher $m$ modes perhaps play a role in the solar case.

%-----------------------------------------------------------------------------------------------------------------------------------

\section{Discussion}

We have investigated the hydrodynamic instability of latitudinal shear during the cooling phase of Type I X-ray bursts, using a shallow water model. We have shown that the geostrophic approximation allows a simple understanding of the r-mode spectrum and criterion for instability, although at the expense of some accuracy in the eigenfunctions near the rotational equator. Considering the simplest shear profile symmetric about the equator, we find that a latitudinal shear of $\gtrsim 2$\%, with the equator rotating more rapidly than the poles, is linearly unstable to hydrodynamic shear instabilities. Instability is related to low frequency buoyancy-driven r-modes that develop a critical layer at which the mode frequency matches the local rotation rate in the differentially rotating shell. At the same time, a modified version of Rayleigh's criterion, based on the potential vorticity, must be satisfied. The unstable modes have low azimuthal wavenumber $m$, e-folding times of roughly a second, and phase speeds close to, but less than, the equatorial rotation rate.

Could the unstable modes be the source of burst oscillations? Identification of the burst oscillations with this instability would be exciting given the importance of understanding angular momentum transport in stratified regions in stellar interiors. The burst oscillations would then give a remarkable opportunity to study angular momentum transport. The requirements for the differential rotation profile are that the equator should be rotating more rapidly than the poles, by about $2$\%. This is very close to the zonal shear during the burst, estimated in \S \ref{sec:zonal} to be $\approx 1$\%. The requirement for instability $s_{\rm crit}>8/3F$ for $F\gg 1$ implies a temperature contrast across the star giving $\Delta H/H\gtrsim 0.3$. Whether this level of temperature contrast is achieved during the burst rise can be addressed by studies of the ignition and propagation of the initial burning front. The fact that the estimated level of differential rotation is close to the stability boundary may explain why burst oscillations are only observed in some bursts. We have considered only one particular differential rotation profile, the simplest possible profile symmetric about the equator. Different profiles may change the stability boundaries and properties of the unstable modes. For example, if ignition occurs off the equator, the differential rotation profile may not be symmetric about the equator. It would be interesting to explore the effect of the ignition site and spreading on the shear profile.

The most important question to answer is why only the $m=1$ mode is excited. Our results suggest that this requires some fine-tuning of the level of differential rotation, and is only possible for spin frequencies close to $300\ {\rm Hz}$. The reason is that we have shown that unlike the two-dimensional case, in which only $l=2$, $m=1$ or $m=2$ modes are unstable (Watson 1981), the shallow water model generally has a range of modes that are unstable. The crucial parameter is $F=(2\Omega R)^2/gH$ (eq.~[\ref{eq:F}]). At the beginning of a burst on a star with rotation frequency $f_s=300\ {\rm Hz}$, $F\approx 100$. In this case, and for a differential rotation of $s\approx 3$\%, just over the threshold for instability, we found that the $m=1$ mode is most unstable. This changes to the $m=2$ mode on increasing the differential rotation to $s\approx 4$\%. For larger values of differential rotation, or for larger $F$, the most unstable modes have $m>2$. This is a problem for the sources with $f_s\approx 600\ {\rm Hz}$, for which $F$ is $4$ times larger ($F\propto f_s^2$). It may be that the non-linear evolution of the instability preferentially drives the $m=1$ amplitude, and in addition Heyl (2004) points out that the higher $m$ modes are much less visible when averaged over the stellar surface. Another possibility is that the $600\ {\rm Hz}$ sources have $m=2$ modes preferentially excited, and therefore have $f_s\approx 300\ {\rm Hz}$. 
However, this may require some fine-tuning of the shear profile, especially so that the excited mode always has the same value of $m$ in a given source.

The change in the nature of the instability with spin frequency is interesting, because it could have implications for the spin distribution of the neutron stars in low mass X-ray binaries. Chakrabarty et al.~(2003) argue that there is an upper limit to the $f_s$ distribution of $\approx 700\ {\rm Hz}$, consistent with observed millisecond radio pulsar spin frequencies, and indicating that some mechanism, perhaps gravitational radiation, prevents accretion from spinning up the neutron star further. If the burst oscillations are connected with shear instabilities, the lack of high frequency oscillations could be because the unstable modes have high $m$ for large $f_s$, and so are unobservable. However, at this point we do not have a clean prediction for the observable spin range.

The lowest burst oscillation frequency is also interesting to consider. EXO~0748-676 has a detected burst oscillation at $45\ {\rm Hz}$ (Villarreal \& Strohmayer 2004), significantly lower than other sources (the next highest burst oscillation frequency is $270\ {\rm Hz}$). For a rotation rate this low, the Rossby deformation radius is comparable to the radius ($F\sim 1$), and so a large amplitude signal is not expected during the burst rise because the rotationally-confined hotspot covers most of the star (Spitkovsky et al.~2002). If the observed oscillation is due to an r-mode, the actual spin frequency of the star could be as much as $\approx 20$\% larger than the observed frequency in EXO~0748-676, since the unstable mode at low $F$ has $l=1$.

We found that the modes driven unstable by the latitudinal shear have $\nu>1$. This makes them more consistent with observed frequency drifts. Heyl (2004) pointed out that although buoyancy-driven r-modes have the required properties to explain burst oscillations (low $m$, a small number of latitudinal nodes, they travel backwards relative to the spin), a problem is that the surface wave (ie.~lowest radial order) associated with the burning layer has a frequency that is too large to explain the observed drifts. We can see the difficulty from equation (\ref{eq:asymptotic}) for the r-mode frequency. In the rotating frame, the frequency of the r-mode can be written
\begin{equation}
m\omega=-{\sqrt{gH}\over R}\left({m\over 2\nu+1}\right)=-24\ {\rm Hz}\ \left({3m\over 2\nu+1}\right){H_3^{1/2}\over R_6}\left({g_{14}\over 2}\right)^{1/2},
\end{equation}
where we scale to $\nu=1$, the mode with the fewest latitudinal nodes. A factor of $3$--$5$ change in temperature as the burst cools gives a frequency drift of $\Delta f\approx 10\ {\rm Hz}$, much larger than observed drifts. A reduced vertical lengthscale $H\approx 100\ {\rm cm}$ rather than $\approx 1000\ {\rm cm}$ is one possible way to reduce the $\nu=1$ frequency to $<10\ {\rm Hz}$, but this requires a mode with more than one radial node. Rescaling to $m=1$ and $\nu=3$, which represents the unstable mode for $F=100$ and $s=0.03$, and using equation (\ref{eq:scaleheight}) for $H$, we find 
\begin{equation}
m\omega=-6.5\ {\rm Hz}\ \left({7m\over 2\nu+1}\right){1\over R_6}\left({T_9\over \mu_m}\right)^{1/2}.
\end{equation}
If we assume that the material burns to $\mu_m\approx 2$ rapidly at the start of the burst, then the frequency drift expected if the layer cools from an initial temperature of $T_9=2$ (as limited by radiation pressure, see Cumming \& Bildsten 2000, eq.~[12]) to $T_9=0.5$ is from $6.5\ {\rm Hz}$ to $3.25\ {\rm Hz}$, in better agreement with observed drifts. This point is complicated by the fact that the frequency drift is likely related to the non-linear development of the instability, since it acts to transport angular momentum within the burning shell, reducing the background shear. The non-linear evolution should be straightforward to follow, by direct numerical integration of the shallow water equations.

Piro \& Bildsten (2005) have recently calculated the spectrum of radial modes for detailed models of the cooling atmosphere following a Type I X-ray burst. They find that the surface wave associated with the burning layer undergoes an avoided crossing with the crustal interface wave as the burst cools. The crustal mode has a temperature-independent frequency, and so this transition terminates the frequency drift, bringing the overall drift into agreement with observed values for $\nu=1$. In addition, this gives a well-defined asymptotic frequency, resolving the question of why the asymptotic frequency observed in bursts should be so stable if it is a mode frequency (say $\approx 3\ {\rm Hz}$ below the spin frequency) rather than the spin frequency itself. Calculations of latitudinal shear instabilities including the full radial structure of the burning layer should be carried out.

With increasing $\nu$, the r-modes should become less visible when the flux is integrated over the stellar surface. Many authors have calculated the lightcurves of rotating neutron stars with hotspots of different sizes and locations, including effects of gravitational light bending (for direct applications to burst oscillations, see Weinberg, Miller, \& Lamb 2001; Nath, Strohmayer, \& Swank 2002; Muno, \"Ozel, \& Chakrabarty 2002; Bhattacharyya et al.~2005). Heyl (2004, 2005) and Lee \& Strohmayer (2005) calculate the expected lightcurves for low order buoyancy-driven r-modes, but for lowest latitudinal order modes only. A calculation of the observable amplitude for the shear unstable modes is needed, especially since these modes have a quite different latitudinal variation to the r-modes, with maximum height perturbation close to the pole rather than the equator. Interestingly, Muno et al.~(2002b) find that a hot spot located close to the pole helps to reduce the harmonic content of the signal.

Oscillations were observed during the superburst in 4U~1636-54 (Strohmayer \& Markwardt 2002). Interestingly, the value of $F$ will be similar for this case. The superburst ignition is believed to occur at depths of $\approx 10^{12}\ {\rm g\ cm^{-2}}$ (Cumming \& Bildsten 2001; Strohmayer \& Brown 2002), where the scale height is $\approx 50\ {\rm m}$. However, taking typical values of $T_9=3$ and $E_F=3\ {\rm MeV}$, the effective lengthscale for the surface wave at these depths is $\approx H (k_BT/E_F)\approx 5\ {\rm m}$, comparable to the scale height in a normal X-ray burst. Therefore, shear instabilities may be relevant for the case of superbursts also. However, the time for burning to spread around the star at such high densities is likely to be much shorter than usual Type I X-ray bursts, so the level of latitudinal differential rotation may be small. In this case, another excitation mechanism for the r-mode needs to be found (such as energy input from nuclear burning, also possibly relevant for normal Type I bursts, see McDermott \& Taam 1987; Strohmayer \& Lee 1996; Lee 2004; Piro \& Bildsten 2004).

Finally, other processes may act to distribute angular momentum during the cooling phase of the burst. For example, a turbulent shear viscosity might act to damp the modes or the background differential rotation. We have ignored the effects of radial shear, and any effect of radial angular momentum transport on the latitudinal shear profile. Cumming \& Bildsten (2000) addressed this question, and were unable to find a robust hydrodynamic angular momentum transport mechanism that acted on a timescale short compared to the burst duration. However, short wavelength baroclinic instabilities should be unstable during the burst tail, and may act to transport angular momentum vertically. Most importantly, they pointed out that even a weak magnetic field could lead to significant angular momentum transport, as it is wound up and acts back on the flow (Spruit 1999). The diffusive magnetorotational instability can also play a role in vertical angular momentum transport within the burning layer (Menou 2004). The strength of the magnetic field in the accreted layers is uncertain for most X-ray bursters, which show no direct evidence of a magnetic field. One possibility is that the magnetic field in the accreted layers is weak due to screening currents in deeper layers (Cumming, Zweibel, \& Bildsten 2001). Two of the burst oscillation sources are also persistent millisecond X-ray  pulsars (SAX J~1808.4-3658, Chakrabarty et al.~2003; XTE~J1814-338, Strohmayer et al.~2003). The burst oscillations from these two sources have unusual properties, such as very rapid frequency drift (Chakrabarty et al.~2003), and significant harmonic component (Strohmayer et al.~2003). The magnetic field apparently has a direct effect on burst oscillations in these two sources. It is possible that a weak magnetic field in the other burst oscillation sources plays a role, and indeed a weak magnetic field is known to significantly change the stability properties of a shearing flow (e.g.~Balbus 1995). The possible effect of a magnetic field on the shearing flows during Type I X-ray bursts should be considered next.

\acknowledgements 
We thank Gil Holder, Anatoly Spitkovsky, and Chris Thompson for useful discussions, and Anna Watts for detailed comments on the paper. We thank Bob Rutledge for useful discussions and for emphasising possible implications for the observed spin distribution of LMXB neutron stars. We acknowledge support from McGill University startup funds, and from the Canadian Institute for Advanced Research.

%-----------------------------------------------------------------------------------------------------------------------------------

\begin{appendix}

\section{Calculation of r-modes using the full shallow water equations}
\label{sec:appa}

In \S \ref{sec:geostrophic}, we made the approximation that the perturbed fluid velocities were in geostrophic balance, and divergence-free. In fact, this assumption must break down near the equator where the horizontal component of the Coriolis force due to horizontal motions vanishes. Here, we calculate the mode spectrum for a rigidly rotating shell using the full shallow water equations, and compare with the results from the geostrophic approximation. 

We consider a uniformly rotating shell with angular speed $\Omega$, and define the parameter\footnote{We follow the notation of Bildsten et al.~(1996). Longuet-Higgins (1968) defines the alternative $\lambda=\omega/2\Omega=1/q$.} $q=2\Omega/\omega$. The numerical solutions of Longuet-Higgins (1968) and Dikpati \& Gilman (1999) start with equations for a stream function $\Psi$ and velocity potential $\Phi$, which measure the curl and divergence of the horizontal velocity. We adopt a shooting method, in which case it is simpler to work with equations for the quantities
\begin{equation}
b=i\left({r\omega\over g}\right)v\sin\theta,
\end{equation}
and the height perturbation $h$ (Longuet-Higgins 1968; Piro \& Bildsten 2004). In terms of these variables, the perturbed momentum and continuity equations may be written
\begin{equation}\label{eq:fulleqnA}
\left({d\over d\mu}-{q\mu  m\over 1-\mu^2}\right)b = h\left[{F\over q^2}-{m^2\over 1-\mu^2}\right]
\end{equation}
and 
\begin{equation}\label{eq:fulleqnB}
\left(1-\mu^2\right){dh\over d\mu}=-b\left(1-q^2\mu^2\right)-q\mu m h.
\end{equation}
We solve these equations using a shooting method, after first factoring out the $(1-\mu^2)^{\left|m\right|/2}$ dependence of the solutions. We start integrating a short distance $\epsilon\ll 1$ away from the pole, and choose the ratio of $h$ and $b$ to ensure that the right hand side of equation (\ref{eq:fulleqnB}) vanishes there. We then look for solutions with definite symmetry about the equator (either $h=0$ or $b=0$ at $\mu=0$).

\begin{figure}
\epsscale{0.6}
\plotone{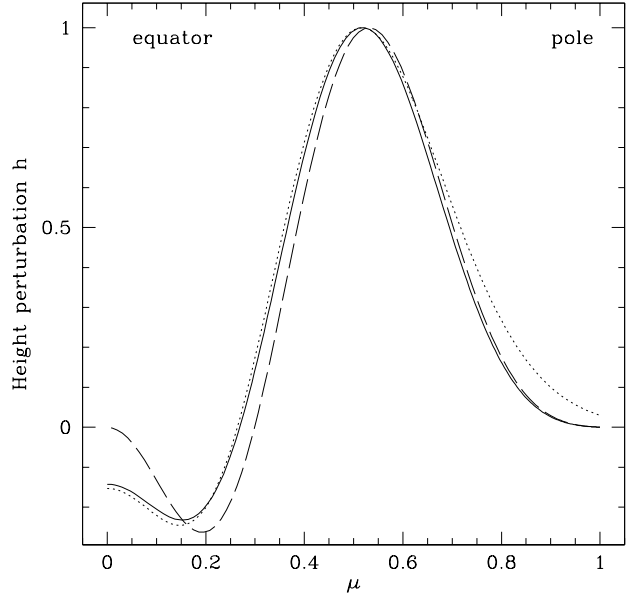}
\caption{Eigenfunctions for modes with $m=1$ and $\nu=3$ in a layer with $F=300$. The solid curve is the solution to the full shallow water equations, the dotted curve is the analytic approximation for large $F$, and the dashed curve is the result of the quasi-geostrophic calculation.\label{fig:mode_compare}}
\end{figure}

Figure \ref{fig:mode_compare} shows the height perturbation for a mode with $\nu=3$, $F=300$. We compare the exact solution calculated here with the analytic asymptotic formula of Longuet-Higgins (1968) and the quasi-geostrophic solution (also, compare with Figure 10 of Longuet-Higgins 1968). The asymptotic expression for $h$ is
\begin{equation}\label{eq:longuetapprox}
h\propto {2\nu+1\over 2m}F^{-1/4}e^{-\eta^2/2}\left[H_{\nu-1}(\eta)+{1\over 2\nu+2}H_{\nu+1}(\eta)\right],
\end{equation}
where $H_\nu(x)$ is the Hermite polymonial of degree $\nu$, and $\eta=F^{1/4}\mu$. There is a good match except for at the equator, as expected. The rotating frame mode frequencies are $-0.01719$ (exact), $-0.01734$ (quasi-geostrophic), and $-2/(2\nu+1)F^{1/2}=-0.01650$ (analytic). The good agreement in the eigenfunctions except at the equator gives us confidence in our quasi-geostrophic results for instability. Note that near the pole, the analytic approximation becomes inaccurate because there is not sufficient room in this case for the eigenfunction to exponentially decay away before it reaches the pole. 

We briefly discuss the relation of our calculations, which average over the vertical structure, to those of BUC96 and Piro \& Bildsten (2004), which include detailed calculations of radial eigenfunctions. Perturbing the momentum and continuity equations (\ref{eq:momentum}) and (\ref{eq:continuity}) and manipulating them into a single equation for the height perturbation $h$, we find
\begin{equation}
{\mathcal L}h = -{F\over q^2}h
\end{equation}
with
\begin{equation}
{\mathcal L}={d\over d\mu}\left[{1-\mu^2\over 1-q^2\mu^2}{d\over d\mu}\right]-{m^2\over (1-\mu^2)(1-q^2\mu^2)}+
{mq(1+q^2\mu^2)\over (1-q^2\mu^2)^2}.
\end{equation}
BUC96 and Piro \& Bildsten (2004) solve the equation ${\mathcal L}h=-\lambda h$, where $\lambda$ measures the horizontal wavevector ($\lambda=l(l+1)$ for slow rotation). In our case, $\lambda=F/q^2=\omega^2R^2/gH$, because in the thin-shell limit only the shallow water surface wave is present. In the slow rotation limit, we recover $\omega^2=l(l+1)gH/R^2=k_\perp^2gH$ as expected.

%-----------------------------------------------------------------------------------------------------------------------------------

\section{Analytic eigenfunctions in the quasi-geostrophic approximation}

We next derive approximate analytic eigenfunctions for the quasi-geostrophic r-modes. Equation (16) with $\Delta\Omega=0$ is
\begin{equation}
{d\over d\mu}\left[\left(1-\mu^2\right){d\psi\over d\mu}\right]-\left[{m^2\over 1-\mu^2}+F\mu^2+q\right]\psi=0.
\end{equation}
where we have defined the parameter $q=2\Omega/\omega$. Now, following Longuet-Higgins (1968), we write $\eta=F^{1/4}\mu$, which gives
\begin{equation}
{d\over d\eta}\left[(1-\mu^2){d\psi\over d\eta}\right]-{m^2\over \sqrt{F}(1-\mu^2)}
+\left(\Lambda-\eta^2\right)=0,
\end{equation}
where $\Lambda=-q/\sqrt{F}=-\sqrt{gH}/R\omega$ (defined so that $\Lambda>0$ since $\omega<0$ for the r-modes). In the limit $F\gg 1$, we take $1-\mu^2=1-\eta^2/\sqrt{F}\approx 1$, and neglect the $m^2$ term, to find
\begin{equation}\label{eq:sho}
{d^2\psi\over d\eta^2}+\left(\Lambda-\eta^2\right)\psi=0.
\end{equation}
The solution is
\begin{equation}
\psi=H_\nu(\eta)e^{-\eta^2/2},
\end{equation}
where $H_\nu(\eta)$ is a Hermite polynomial, and $\Lambda=2\nu+1$. This gives the expected mode frequencies, since $\omega=-\sqrt{gH}/R\Lambda$.

The height perturbation in the geostrophic approximation is $\propto \mu\psi$ (eq.~[\ref{eq:heightgeo}]), giving
\begin{equation}
h\propto \eta F^{-1/4} H_\nu(\eta)e^{-\eta^2/2}.
\end{equation}
Using the recurrence relation for Hermite polynomials, $H_{n+1}(x)=2xH_n(x)-2nH_{n-1}(x)$, this can be rewritten
\begin{equation}\label{eq:myapprox}
h\propto F^{-1/4} e^{-\eta^2/2}\left[H_{\nu-1}(\eta)+{1\over 2\nu}H_{\nu+1}(\eta)\right].
\end{equation}
which differs from the approximation to the full shallow water equations derived by Longuet-Higgins (1968) (eq.~[\ref{eq:longuetapprox}]) by a factor of $2\nu$ rather than $2\nu+2$ in the denominator of the second term.

Finally, note that equation (\ref{eq:sho}) is the same as the equation describing a simple harmonic oscillator in quantum mechanics. We can now understand the behavior of the r-mode eigenfunctions for increasing $\nu$. The eigenfunction is able to propagate to the classical turning point at $\eta=\sqrt{\Lambda}$, or $\mu=\sqrt{2\nu+1}/F^{1/4}=(2\nu+1)q^{-1/2}$, and then evanesces towards the pole. 

\end{appendix}

%-----------------------------------------------------------------------------------------------------------------------------------


\begin{references}

\noindent
Balbus, S.~A.\ 1995, \apj, 453, 380 

\noindent
Bhattacharya, D. 1995, in X-Ray Binaries, ed. W. H. G. Lewin, J. van Paradijs, \& E. P. J. van den Heuvel (Cambridge: Cambridge University Press), 233

\noindent
Bhattacharyya, S., Strohmayer, T.~E., Miller, M.~C., \& Markwardt, C.~B.\ 2005, \apj, 619, 483 

\noindent
Bildsten, L., Ushomirsky, G., \& Cutler, C.\ 1996, \apj, 460, 827 

\noindent
Chakrabarty, D., Morgan, E.~H., Muno, M.~P., Galloway, D.~K., Wijnands, R., van der Klis, 
M., \& Markwardt, C.~B.\ 2003, \nat, 424, 42 

\noindent
Charbonneau, P., Dikpati, M., \& Gilman, P., 1999, \apj, 528, 523

\noindent
Cumming, A., \& Bildsten, L.\ 2000, \apj, 544, 453

\noindent
Cumming, A., \& Bildsten, L.\ 2001, \apj, 559, L127

\noindent
Cumming, A., Morsink, S.~M., Bildsten, L., Friedman, J.~L., \& Holz, D.~E.\ 2002, \apj, 564, 343

\noindent
Cumming, A., Zweibel, E., \& Bildsten, L.\ 2001, \apj, 557, 958 

\noindent
Dikpati, M., \& Gilman, P.~A. 1999, \apj, 512, 417

\noindent
Drazin, P.~G., \& Reid, W.~H. 1981, "Hydrodynamic Stability" (Cambridge: Cambridge University Press)

\noindent
Dziembowski, W., \& Kosovichev, A. 1987, Acta Astron., 37, 341

\noindent
Franco, L.~M.\ 2001, \apj, 554, 340 

\noindent
Garaud, P. 2001, \mnras, 324, 68

\noindent
Gilman, P.~A., \& Fox, P.~A. 1999, \apj, 510, 1018

\noindent
Heyl, J.~S.\ 2004, \apj, 600, 939 

\noindent
Heyl, J.~S.\ 2005, \mnras, submitted (astro-ph/0502518)

\noindent
Houghton, J.~T. 1986, ``The Physics of Atmospheres'', 2nd edition (Cambridge: Cambridge University Press)

\noindent
Kaaret, P., Zand, J.~J.~M.~i., Heise, J., \& Tomsick, J.~A.\ 2003, \apj, 598, 481 

\noindent
Lee, U.\ 2004, \apj, 600, 914 

\noindent
Lee, U., \& Strohmayer, T.~E.\ 2005, \mnras, submitted (astro-ph/0502502)

\noindent
Lewin, W.~H.~G., van Paradijs, J., \& Taam, R.~E.\ 1993, Space Science Reviews, 62, 223 

\noindent
Lewin, W. H. G., van Paradijs, J., \& Taam, R. E. 1995, in X-Ray Binaries,
ed. W. H. G. Lewin, J. van Paradijs, \& E. P. J. van den Heuvel (Cambridge:
CUP), 175

\noindent
Longuet-Higgins, M.~S. 1968, Phil.~Trans.~Royal~Soc.~London, Series A, 262, 1132

\noindent
McDermott, P.~N., \& Taam, R.~E.\ 1987, \apj, 318, 278 

\noindent
Menou, K. 2004, \mnras, 352, 1381

\noindent
Muno, M.~P., Chakrabarty, D., Galloway, D.~K., \& Psaltis, D.\ 2002a, \apj, 580, 1048 

\noindent
Muno, M.~P., Fox, D.~W., Morgan, E.~H., \& Bildsten, L.\ 2000, \apj, 542, 1016 

\noindent
Muno, M.~P., Galloway, D.~K., \& Chakrabarty, D.\ 2004, \apj, 608, 930 

\noindent
Muno, M.~P., {\" O}zel, F., \& Chakrabarty, D.\ 2002b, \apj, 581, 550 

\noindent
Muno, M.~P., {\" O}zel, F., \& Chakrabarty, D.\ 2003, \apj, 595, 1066 

\noindent
Nath, N.~R., Strohmayer, T.~E., \& Swank, J.~H.\ 2002, \apj, 564, 353 

\noindent
Papaloizou, J.~C.~B., \& Pringle, J.~E. 1985, \mnras, 213, 799

\noindent
Pedlosky, J. 1987, ``Geophysical Fluid Dynamics'', 2nd edition (New York: Springer-Verlag)

\noindent
Piro, A.~L., \& Bildsten, L.\ 2004, \apj, 603, 252 

\noindent
Piro, A.~L., \& Bildsten, L.\ 2005, \apj, submitted (astro-ph/0502546)

\noindent
Spitkovsky, A., Levin, Y., \& Ushomirsky, G.\ 2002, \apj, 566, 1018

\noindent
Spruit, H.~C. 1999, \aap, 349, 189

\noindent
Strohmayer, T. E., \& Bildsten, L. 2003, in Compact Stellar X-Ray
Sources, eds. W.H.G. Lewin and M. van der Klis (Cambridge: Cambridge
University Press) (astro-ph/0301544)

\noindent
Strohmayer, T.~E., \& Brown, E.~F.\ 2002, \apj, 566, 1045 

\noindent
Strohmayer, T.~E., Jahoda, K., Giles, A.~B., \& Lee, U.\ 1997a, \apj, 486, 355 

\noindent
Strohmayer, T.~E., \& Lee, U.\ 1996, \apj, 467, 773

\noindent
Strohmayer, T.~E., \& Markwardt, C.~B.\ 1999, \apjl, 516, L81 

\noindent
Strohmayer, T.~E., \& Markwardt, C.~B.\ 2002, \apj, 577, 337 

\noindent
Strohmayer, T.~E., Zhang, W., \& Swank, J.~H.\ 1997b, \apjl, 487, L77 

\noindent
Strohmayer, T.~E., Markwardt, C.~B., Swank, J.~H., \& in't Zand, J.\ 2003, \apjl, 596, L67 

\noindent
Strohmayer, T.~E., Zhang, W., Swank, J.~H., Smale, A., Titarchuk, L., Day, C., \& Lee, U.\ 
1996, \apjl, 469, L9

\noindent
Strohmayer, T.~E., Zhang, W., Swank, J.~H., White, N.~E., \& Lapidus, I.\ 1998a, \apjl, 498, 
L135 

\noindent
Strohmayer, T.~E., Zhang, W., Swank, J.~H., \& Lapidus, I.\ 1998b, \apjl, 503, L147 

\noindent
van Straaten, S., van der Klis, M., Kuulkers, E., \& M{\' e}ndez, M.\ 2001, \apj, 551, 907 

\noindent
Villarreal, A.~R., \& Strohmayer, T.~E.\ 2004, \apjl, 614, L121 

\noindent
Watson, M. 1981, Geophys.~Astrophys.~Fluid Dynamics, 16, 285

\noindent
Watts, A.~L., Andersson, N., Beyer, H., \& Schutz, B.~F.\ 2003, \mnras, 342, 1156 

\noindent
Watts, A.~L., Andersson, N., \& Williams, R.~L.\ 2004, \mnras, 350, 927 

\noindent
Weinberg, N., Miller, M.~C., \& Lamb, D.~Q. 2001, \apj, 546, 1098

\end{references}
\end{document}